\documentclass{article}

\usepackage{arxiv}

\usepackage[utf8]{inputenc} 
\usepackage[T1]{fontenc}    
\usepackage{hyperref}       
\usepackage{url}            
\usepackage{booktabs}       
\usepackage{amsfonts}       
\usepackage{nicefrac}       
\usepackage{microtype}      
\usepackage{lipsum}
\usepackage{graphicx}
\usepackage{abstract}
\usepackage{amsmath,amsthm,amssymb,amsfonts}
\graphicspath{ {./images/} }
\usepackage{caption}

\title{SwG-former: A Sliding-Window Graph Convolutional Network for Simultaneous Spatial-Temporal Information Extraction in Sound Event Localization and Detection}

\author{
 Weiming Huang \\
  School of Computer Engineering and Science\\
   Shanghai University\\
  \texttt{nerv.wm.huang@outlook.com} \\
   \And
 Qinghua Huang* \\
  School of Communication and Information Engineering\\
  Shanghai University\\
  \texttt{qinghua@shu.edu.cn} \\
  \And
 Liyan Ma \\
  School of Computer Engineering and Science\\
  Shanghai University\\
  \texttt{liyanma@shu.edu.cn} \\
  \And
  \And
   Chuan Wang \\
  School of Communication and Information Engineering\\
  Shanghai University\\
  \texttt{wangchuan1101@shu.edu.cn} \\
}

\begin{document}
\maketitle
\captionsetup[figure]{labelfont={bf},labelformat={default},labelsep=period,name={Fig.}}

\begin{abstract}
Sound event localization and detection (SELD) involves sound event detection (SED) and direction of arrival (DoA) estimation tasks. SED mainly relies on temporal dependencies to distinguish different sound classes, while DoA estimation depends on spatial correlations to estimate source directions. This paper addresses the need to simultaneously extract spatial-temporal information in audio signals to improve SELD performance. A novel block, the sliding-window graph-former (SwG-former), is designed to learn temporal context information of sound events based on their spatial correlations. The SwG-former block transforms audio signals into a graph representation and constructs graph vertices to capture higher abstraction levels for spatial correlations. It uses different-sized sliding windows to adapt various sound event durations and aggregates temporal features with similar spatial information while incorporating multi-head self-attention (MHSA) to model global information. Furthermore, as the cornerstone of message passing, a robust Conv2dAgg function is proposed and embedded into the block to aggregate the features of neighbor vertices. As a result, a SwG-former model, which stacks the SwG-former blocks, demonstrates superior performance compared to recent advanced SELD models. The SwG-former block is also integrated into the event-independent network version 2 (EINV2), called SwG-EINV2, which surpasses the state-of-the-art (SOTA) methods under the same acoustic environment.
\end{abstract}


\section{Introduction}
\setlength{\parindent}{2em}

In the presence of overlapping auditory signals, the human ear has an innate ability to effortlessly localize each sound source and discern the sound of interest. How can machines have a similar capability to localize and distinguish overlapping sounds? Sound event localization and detection (SELD) aims to tackle this challenge and has extensive applications such as monitoring systems \cite{ref1}, smart homes \cite{ref2}, wildlife protection \cite{ref3}, and intelligent conference rooms \cite{ref4}. 

SELD is designed to identify the classes, onset, and offset of sound events from multichannel audio signals and estimate the spatial localizations of the corresponding sound events, which integrates the subtasks of sound event detection (SED) and direction of arrival (DoA) estimation. Before deep learning, many SED methods, including the Gaussian mixture model-hidden Markov model (GMM-HMM) \cite{ref5} and non-negative matrix factorization (NMF) \cite{ref6}, were used for modeling sound events and separating sound sources. At the same time, DoA estimation could be completed well by beamformers, such as the minimum variance distortionless response (MVDR) \cite{ref7} and steered-response power (SRP) \cite{ref8} in an anechoic acoustic environment. However, these methods could not perform robustly in complex acoustic environments, such as interference, overlapping sources, and reverberation. To overcome these challenges and jointly optimize two subtasks, the SELD system should simultaneously capture spatial correlations and model temporal dependencies. This is because SED mainly relies on temporal dependencies to distinguish different sound classes, and DoA estimation depends on spatial correlations recorded in amplitude and phase differences between microphones to estimate source directions. 

Based on deep neural network (DNN), the SELDnet \cite{ref9} divides the SELD task into several parts: feature pre-extraction, network model, and output format. The input features for SELDnet consist of multichannel log-Mel spectrograms concatenated along with the intensive vectors (IVs) processed by the feature extractor. SELDnet employs a convolutional recurrent neural network (CRNN) as the backbone of the network, which exploits three convolutional blocks for high-level features extraction and recurrent neural networks (RNNs) based on the bidirectional gated recurrent unit (BiGRU) for temporal context modeling. Then, the output of RNN is fed into two parallel, fully connected (FC) layers to predict SED classification and DOA regression, respectively. To detect overlapping sound events with the same class but different DOAs, the event-independent network v2 (EINV2) \cite{ref10} improves SELD output in the track-wise format. The Conv-Conformer \cite{ref11} is an extension of EINV2 using Conformer blocks \cite{ref12} and dense blocks to model sequence context information. Instead of outputting in separated branches, CRNN \cite{ref13} employs the activity-coupled Cartesian direction of arrival (ACCDOA) output format to unify SED and DoA into one loss function and regard the SELD task as a cartesian regression task. 

However, these methods employ CNN to extract spatial information and BiGRU or Conformer to extract temporal features separately. In this paper, the interdependence of spatial-temporal information in audio signals is exploited for simultaneous extraction to enhance the model performance, as experimentally discussed in section 4.4.

In contrast, to achieve more flexible audio representation processing compared to the conventional spectrogram approach, sliding-window graph-former (SwG-former) and SwG-EINV2 models were proposed for better SELD performance, based on CRNN and EINV2 frameworks, respectively. As the cornerstone of these models, the SwG-former block leverages a novel graph representation to extract spatial-temporal information simultaneously for joint optimization of SED and DoA tasks. The SwG-former block exploits graph convolutional network (GCN) in non-Euclidean space to learn temporal context information of sound events based on their spatial correlations while incorporating multi-head self-attention (MHSA) \cite{ref14} to model global information. Specifically, it uses different-sized sliding-windows to robustly adapt instantaneous or long-duration dynamic sound events and aggregates temporal features of sound events with similar spatial information. A more robust convolution 2D aggregation (Conv2dAgg) function and FC layer are embedded into the SwG-former block to aggregate these features and update inner temporal features, respectively. Experimental results show better performance for SwG-former model than recent advanced SELD models, which have simple architecture and could be seamlessly integrated into other models. Further, SwG-EINV2 achieves a state-of-the-art (SOTA) performance under the same acoustic environment.

The main contributions of the paper can be summarized as follows:
\begin{enumerate}
\item A universal and novel method is proposed to convert audio signals into graph data structures, efficiently extracting spatial-temporal features concurrently. It is the first study that employs GCN for the SELD task.
\item A plug-and-play SwG-former block is proposed to learn temporal context information at different levels of resolutions and dynamically construct graph vertices to capture spatial correlations. A more robust Conv2dAgg function embedded into the block has stronger fitting capabilities to pass messages among vertices than the standard graph aggregation function.
\item A simple and efficient SwG-former model, stacking the SwG-former blocks, is designed for the SELD task and displays superior results compared with other methods. Further, it is integrated with the EINV2 framework and surpasses the SOTA methods.
\end{enumerate}

The remainder of the paper is organized as follows: Section 2 introduces the theoretical background of GCN. Section 3 describes the designed SwG-former block, the proposed Conv2dAgg function, and the architectures of SwG-former and SwG-EINV2 models. The dataset, experimental setup, and a series of ablation studies aimed at exploring superior model performance, along with the experimental results and a visualization analysis, are presented in Section 4. Finally, the conclusion is given in Section 5.

\section{Related work}
\subsection{Existing SELD methods}

Significant progress for SELD has been achieved following a series of challenges, including the detection and classification of acoustic scenes and events (DCASE) challenge \cite{ref9} and the IEEE ICASSP grand challenge-L3DAS \cite{ref15}. Emerging models can be primarily classified into multi-branch and single-branch output formats.
Adavanne et al. \cite{ref9} proposed the SELDNet implemented by a CRNN and outputs two branches to predict SED and DoA separately. Cao et al. \cite{ref10} proposed the EINV2 based on the soft-parameter sharing mechanism of CNN blocks and MHSA. The EINV2 further adopts a track-wise output format for detecting overlapping sound events of the same class but with different DoAs. Hu et al. \cite{ref11} improved EINV2 by incorporating Conformer and dense blocks. However, the model is parameter-heavy due to the integration of different models.

Shimada et al. \cite{ref13} regarded SELD as a Cartesian regression task, merging SED and DoA representations into a single-branch output called ACCDOA. This method overcomes the loss balance issue in dual-branch output. Later, Shimada et al. \cite{ref16} introduced a multi-ACCDOA format that handles overlapping instances from the same class while maintaining a class-wise output format. Consequently, multi-ACCDOA became a prevalent output format. Wu et al. \cite{ref17} enhanced the original CRNN by inserting a shallow multi-layer perceptron-mixer (MLP-Mixer) between the convolution filters and the recurrent layers to model inter-channel audio patterns intricately. Wang et al. \cite{ref18} introduced a model that combines CRNN, MHSA technique, and a CNN-Transformer encoder to explore various model structures incorporating attention mechanisms. Wang et al. \cite{ref19,ref20} used a ResNet-Conformer structure as the backbone network, where ResNet is used to extract spatial features, and Conformer is adopted to mode temporal context dependencies. Shul et al. \cite{ref21} proposed the divided spectro-temporal attention model to construct sequential contextual relationships in the frequency and temporal domains separately. However, only some of these methods could simultaneously exploit spatial-temporal patterns to better jointly optimize SED and DoA tasks.

\subsection{Graph convolutional networks}
To overcome the challenge posed by the inability of CNN to process non-Euclidean data, graph neural networks (GNNs) emerged. Early GNNs \cite{ref22} were primarily applied to address strict graph theory problems on graphs $\mathsf{\mathcal{G}}=(V,\mathsf{\mathcal{E}})$. The $\mathsf{\mathcal{G}}$ is composed of vertices set $V$ and edges set $\mathsf{\mathcal{E}}$. The GNNs were usually applied to social networks \cite{ref23}, traffic forecasting \cite{ref24,ref25}, chemical molecular structure prediction \cite{ref26,ref27}, recommendation engines \cite{ref28}, and so on. It was not until Bruna et al. \cite{ref29} first proposed the spectral GCN and spatial GCN that their application in audio-related tasks became prosperous. Shirian et al. \cite{ref30} proposed compact graph convolutional networks for speech emotion recognition tasks. Tzirakis et al. \cite{ref31} treated audio channels as vertices to construct graph structures and adopted GCN to extract the spatial correlation among different channels for multi-channel speech enhancement. Wang et al. \cite{ref32} encoded more structural details based on GCN for speech separation.

The implementation of GCN on audio-related tasks is mainly based on the spectral GCN, which decomposes the graph Laplacian matrix to aggregate neighbor vertex features. However, the operation of eigenvalue decomposition will bring unbearable costs once the graph structures become more complex. Based on this, the spatial GCN method aggregates information from neighbor vertices and avoids the eigenvalue decomposition. Hence, the spatial GCN is applied in the SELD task. 

\subsection{Aggregation functions of spatial GCNs}

The aggregation function is proposed for spatial GCN to aggregate the vertex features from neighbor vertices. It can be mainly divided into parameter-less max \cite{ref33,ref34,ref35}, mean  \cite{ref36}functions, and parameterized functions like long short-term memory (LSTM). Hamilton et al. \cite{ref34} empirically found that max and LSTM outperform mean aggregation. Veličković et al. \cite{ref37}proposed graph attention networks (GATs) that learn the attention weights between the central vertex and the neighbor vertices through an attention mechanism \cite{ref38}. Furthermore, Xu et al. \cite{ref39} proposed graph isomorphism networks (GIN) that adopt sum aggregation, which demonstrates discriminative power equivalent to the Weisfeiler-Lehman graph isomorphism test. Additionally, Li et al. \cite{ref40} adopted max-relative (MR) graph convolution to aggregate neighbor vertex features through max pooling operation without learnable parameters. These aggregation functions can handle variable-length data but at the expense of some fitting ability, which means each neighbor vertex aggregated by the same weight $w$, defined as:

\begin{flalign}\label{eq1}
&&
{{\operatorname{Agg}}_{1}}\left( {{\mathbf{h}}_{i}} \right)=\sum\limits_{{{\mathbf{h}}_{j}}\in \mathsf{\mathcal{N}}\left( {{\mathbf{h}}_{i}} \right)}{w{{\mathbf{h}}_{j}}},
&&
\end{flalign}
where ${{\mathbf{h}}_{j}}\in \mathsf{\mathcal{N}}({{\mathbf{h}}_{i}})$ is the feature of  the vertex ${{v}_{i}}$ and ${{\mathbf{h}}_{j}}\in \mathsf{\mathcal{N}}({{\mathbf{h}}_{i}})$ $j\in 1,...,k$ is neighbor vertex features of ${{v}_{i}}$. In contrast, the CNN aggregation function is denoted as:

\begin{flalign}\label{eq2}
&&
    {{\operatorname{Agg}}_{2}}\left( {{\mathbf{h}}_{i}} \right)=\sum\limits_{{{\mathbf{h}}_{j}}\in \mathsf{\mathcal{N}}\left( {{\mathbf{h}}_{i}} \right)}{{{w}_{j}}{{\mathbf{h}}_{j}}},	
&&
\end{flalign}
where there are $k$ weights ${{w}_{i}}$ ($j\in 1,...,k$) for the convolutional kernel. As a result, CNN has stronger fitting capabilities than GCN but at the expense of handling variable-length data. In the proposed GCN aggregation function, information from these neighbor vertices can be aggregated by Eq. (\ref{eq2}) because each vertex is associated with a fixed number of neighbor vertices.

\section{Proposed SwG framework}
\begin{figure}[ hb] 
    \centering
    \includegraphics[width=1.0\textwidth]{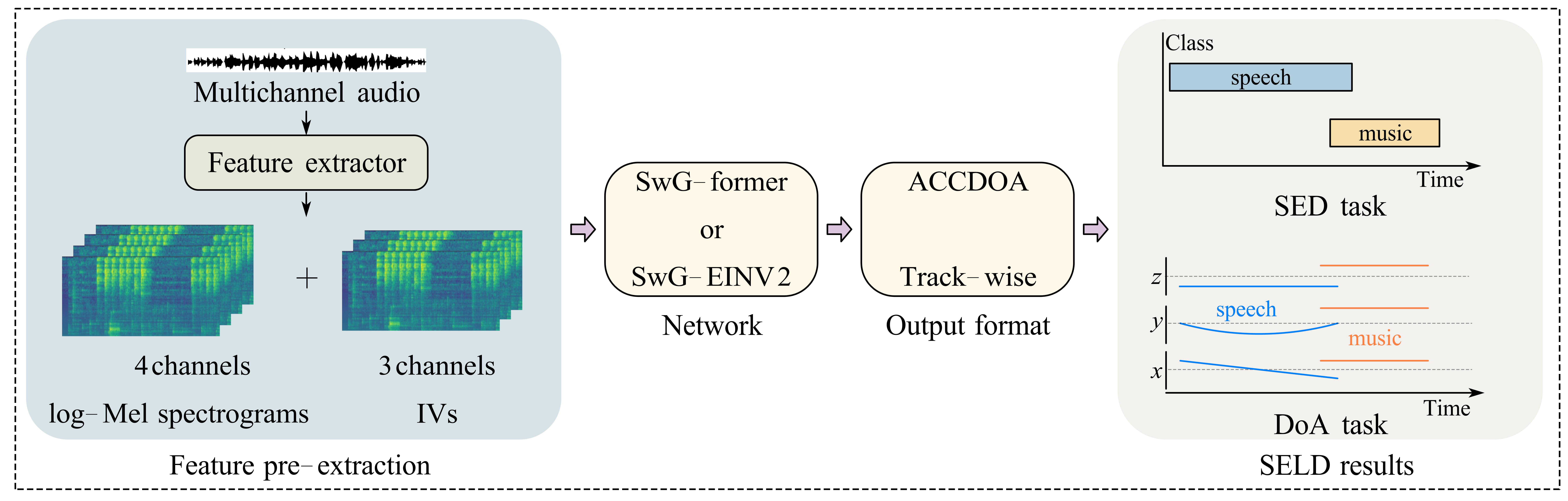}
  
    \caption{Flowchart of SwG framework.}

  \label{fig:fig1}
\end{figure}

Given a multichannel audio signal, the SELD task aims to identify classes, onset, and offset of sound events and estimate the spatial localizations simultaneously. Firstly, multichannel audio is processed to log-Mel spectrograms with IVs and fed into SwG-former or SwG-EINV2 network models. Then, the network models output SELD results in ACCDOA \cite{ref13} and Track-wise \cite{ref11} formats, respectively. The overall flow of the SwG framework is illustrated in Fig. \ref{fig:fig1}.

\subsection{SwG-former block}
\begin{figure}[ hb] 
    \centering
    \includegraphics[width=0.73\textwidth]{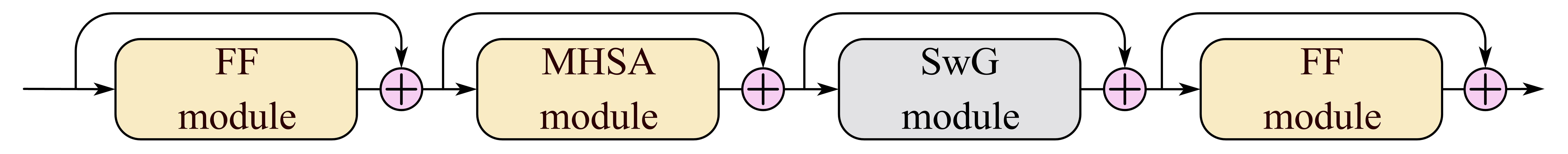}
  
    \caption{Proposed SwG-former block. }

  \label{FIG:fig2}
\end{figure}

The key that jointly optimizes SED and DoA tasks is to capture spatial-temporal information simultaneously. The SwG-former block is capable of processing audio signals by extracting spatial correlations while concurrently modeling both local and global temporal dependencies. The SwG-former block retains a pair of macaron-like Feed Forward (FF) modules, sandwiching other modules from the Conformer block. It replaces the convolution module with the SwG module, which captures higher abstraction levels for spatial information compared with the convolution module. The SwG module learns the spatial correlations inherent in dynamic sound scenes by dynamically changing graph structures at different layers and extracts local temporal dependencies. The MHSA module \cite{ref12} assists in modeling global context dependencies in the audio sequence. Fig. \ref{FIG:fig2} shows the architecture of the SwG-former block.

\subsubsection{SwG module}
\begin{figure}[ hb] 
    \centering
    \includegraphics[width=1\textwidth]{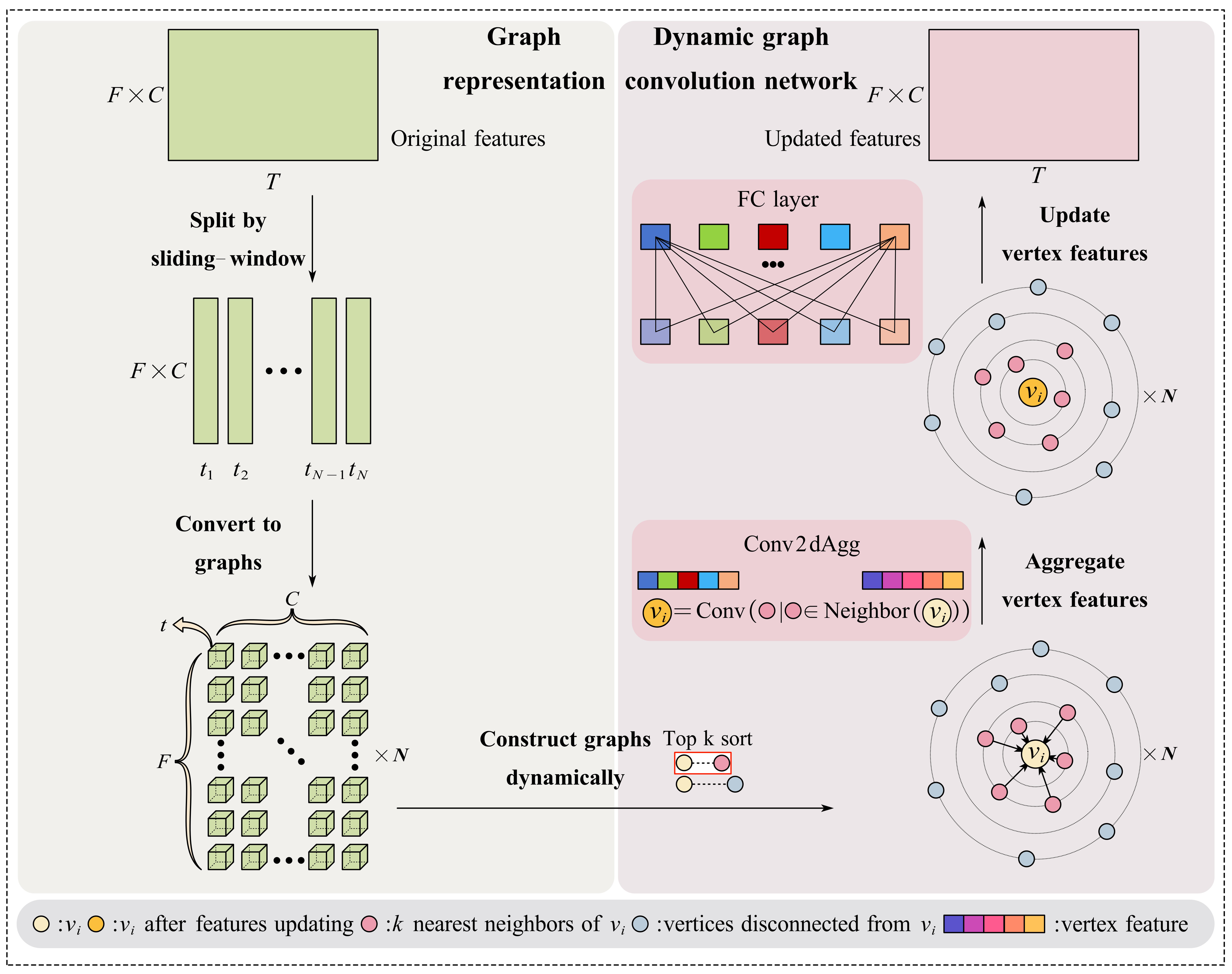}
  
    \caption{Schematic illustration of the SwG module. The original features are converted to $N$ graphs, and vertices are dynamically constructed in the F-C domain within each graph to extract spatial correlations of audio features. These vertex features are aggregated and updated by the proposed Conv2dAgg function and FC layer for more diverse features. As a result, the features of different class sound events are more distinctive for better model performance. $T, F$, and $C$ represent the time frame, frequency bin, and channel, respectively. }

  \label{FIG:fig3}
\end{figure}

The proposed SwG module aggregates the temporal information of sound events with similar spatial correlations to diversify the features of different sound events. This interdependence of spatial-temporal information is exploited for simultaneous extraction to enhance the model performance. For robustly adapting various durations of sound events, the SwG module employs different-sized sliding windows to split original features at different levels of resolutions. To extract spatial correlations of sound events, the SwG module dynamically constructs graph vertices and connections in the frequency-channel (F-C) domain based on the similarities of vertex features. 
The SwG module contains two processes: graph representation aims to transform the audio signal into a graph representation, and dynamic graph convolution network supports message passing among vertices. Additionally, Fig. \ref{FIG:fig3} provides a schematic representation of how the SwG module learns temporal context information at different levels of resolutions and extracts F-C features to capture spatial correlations: firstly, the spectrogram representation is converted into a graph representation with a non-Euclidean structure; vertex vi constructs connections with its k nearest neighbors during the dynamic graph construction process; vi aggregates and updates features from its neighbor vertices by Conv2dAgg function and FC layer; subsequently, the generated N graphs are concatenated along the time axis into a spectrogram structure and fed into the next SwG module with other-sized windows or FF module. 

\subsubsection{Proposed graph representation of audio signals}

Let $\mathsf{\mathcal{G}}=(V,\mathsf{\mathcal{E}})$ denote a graph where $V$ and ${\mathcal{E}} $ denote the vertices and edges. The input feature $\mathbf{x}' \in {{\mathbb{R}}^{T\times F\times C}}$ is split into   non-overlapping equal chunks using a sliding-window along with the time dimension, where the sliding-window size is $t$ and $ t = T / N$. Each chunk with the size of $t\times F\times C$ is denoted as a graph and each graph possesses the vertices $V=\left\{ v_1,...,v_n \right\} $ where $n=F\times C$. Then the feature vectors of vertices are denoted as $\mathbf{H}=\left[ \mathbf{h}_1,...,\mathbf{h}_{\mathbf{n}} \right]$where ${{\mathbf{h}}_{i}}\in {{\mathbb{R}}^{t}}$. According to the similarity of each pair of vertices, each vertex ${{v}_{i}}$ finds its $k$ nearest neighbors. The time chunks are fed into the SwG module to extract latent spatial-temporal features. In the shallower layers, $t$ is set as 5 to extract instantaneous temporal features. In the deeper layers, $t$ is set as 25 to obtain long-duration temporal features.

The proposed graph representation for audio signals shows several compelling advantages. Primarily, the graph offers a generalized data structure, where the sequence of an audio signal can be interpreted as a particular case of a graph. With its non-Euclidean structure, this graph structure extends more flexibility than the spectrogram representation. It allows for the simultaneous extraction of latent spatial-temporal features based on the sliding-windows to learn temporal context information at different levels of granularity. Furthermore, by dynamically constructing graph vertices within the F-C domain, the graph representation learns the spatial correlations inherent in dynamic sound scenes. Furthermore, the proposed graph representation can be effectively leveraged to tackle various audio-related tasks.

\subsubsection{ Dynamic graph convolution network}
Most GCNs come with a predefined topological structure, where vertex features are updated in each layer with a fixed graph structure. That could not better adapt to the characteristics of different data. This paper uses dynamic graph convolution to extract feature information better. The dynamic GCN constructs a weight matrix of edge sets dynamically in each layer by calculating the similarity of each pair of vertex features in the current feature space. Each vertex constructs connections by the k-nearest neighbors (KNN) algorithm to aggregate and update vertex information. 

For the overall graph, the dynamic graph convolution network can be formulated as follows: 
\begin{flalign}\label{eq3}
&&
    \mathbf{{G}'}=\operatorname{Update}\left( \mathbf{G},\operatorname{Aggregate}\left( \mathbf{G},{{\mathbf{W}}_{\operatorname{agg}}} \right),{{\mathbf{W}}_{\operatorname{update}}} \right),
&&
\end{flalign}
where $\mathbf{G}$ and $\mathbf{{G}'}$ are the input and output of the dynamic graph convolution network, ${{\mathbf{W}}_{\operatorname{agg}}}$ and ${{\mathbf{W}}_{\operatorname{update}}}$, as the core of GCN, are the learnable weights of the $\operatorname{Aggregate}\left( \cdot  \right)$ and $\operatorname{Update}\left( \cdot  \right)$ operations, respectively.

For each vertex, the $\operatorname{Aggregate}\left( \cdot  \right)$ operation corresponds to the aggregation function, denoted as $\operatorname{g}\left( \cdot  \right)$. This function calculates the vertex representation by aggregating the features of neighboring vertices. The $\operatorname{Update}\left( \cdot  \right)$ operation corresponds to the update function, symbolized as $\operatorname{f}\left( \cdot  \right)$. This function is employed to compute the new vertex representation derived from the aggregated information. Equation (\ref{eq3}) can be expressed as as follows:
\begin{flalign} \label{4}
&&
    {{\mathbf{{h}'}}_{i}}=\operatorname{f}\left( {{\mathbf{h}}_{i}},\operatorname{g}\left( {{\mathbf{h}}_{i}},\mathsf{\mathcal{N}}\left( {{\mathbf{h}}_{i}} \right),{{\mathbf{W}}_{\operatorname{agg}}} \right),{{\mathbf{W}}_{\operatorname{update}}} \right),	
&&
\end{flalign}	
where $\mathsf{\mathcal{N}}\left( {{\mathbf{h}}_{i}} \right)$ is neighbor vertex features of ${{v}_{i}}$.

In the proposed graph structure, each vertex is associated with a fixed number of $k$ neighbor vertices instead of variable-length graph data. As a result, information from these neighbor vertices can be aggregated using $1\times k$ convolutional kernels (Eq. (\ref{eq2})). The proposed convolution 2D aggregation (Conv2dAgg) and feature update functions are defined as follows:
\begin{flalign} \label{5}
&&
    \operatorname{g}\left( \cdot  \right)=\operatorname{Conv}\left( {{\mathbf{h}}_{j}}|{{\mathbf{h}}_{j}}\in \mathsf{\mathcal{N}}\left( {{\mathbf{h}}_{i}} \right) \right),
&&
\end{flalign}
\begin{flalign} \label{6}
&&
    \operatorname{f}\left( \cdot  \right)=\operatorname{MLP}\left( {{\mathbf{h}}_{i}},\operatorname{g}\left( {{\mathbf{h}}_{i}} \right),{{\mathbf{W}}_{\operatorname{update}}} \right),	
&&
\end{flalign}
where Conv in aggregation function is 2D convolution, the vertex feature updater f is a multilayer perceptron (MLP) with the batch normalization and Gaussian error linear unit (GeLU) \cite{ref41} as the activation function.  

In order to facilitate the interaction of temporal features of each vertex, the transform model is employed behind the dynamic graph convolution network (Eq. (\ref{eq3})) and denoted as:
\begin{flalign} \label{8}
&&
    \mathbf{Y}=\operatorname{GeLU}\left( \mathbf{{G}'}{{\mathbf{W}}_{1}} \right){{\mathbf{W}}_{2}}+\mathbf{{G}'}
&&
\end{flalign}
where $\mathbf{Y}$ is the output of the transform model and also the input of the following dynamic graph convolution network layer, ${{\mathbf{W}}_{1}}$ and ${{\mathbf{W}}_{2}}$ are the weights of the FC layers. The FC layers increase the diversity of vertex features and alleviate the over-smoothing problem in deep GCN \cite{ref42}. 

\subsection{Network architectures}
This section describes SwG-former, a simple and efficient model. It processes SED and DoA tasks within a single branch and outputs in the ACCDOA format. To have an extensive comparison with complex models that output in the track-wise format, the SwG-former blocks take the place of the Conformer blocks within the EINV2 framework, named SwG-EINV2. The seamless integration of SwG-former block substantiates its generality.
\subsubsection{Input features}
\begin{figure}[ht] 
    \centering
    \includegraphics[width=0.48\textwidth]{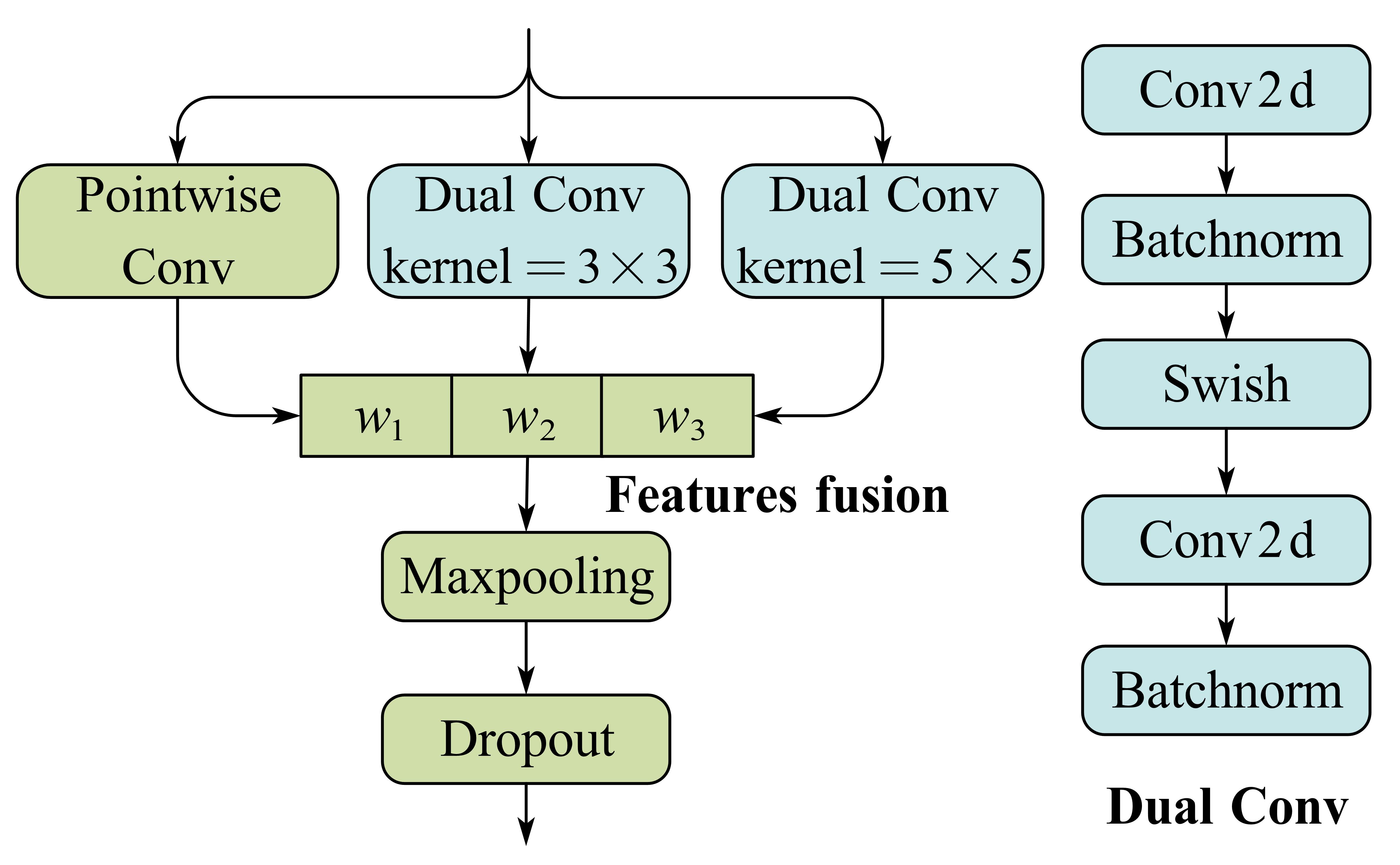}

    \caption{MS-Conv block.}
  \label{FIG:fig4}
\end{figure}

Most SELD models stack CNNs with the same kernels to facilitate the extraction of feature representations. To enhance the diversity of feature embeddings, the input features undergo processing through multi-scale onvolution (MS-Conv) blocks, accompanied by a dimensionality reduction to alleviate computational load, as delineated in Fig. \ref{FIG:fig4}. The first-order Ambisonics (FOA) format 4-channel audio dataset is employed and extracted to 4-channel log-Mel spectrograms and with 3-channel IVs concatenated along with channel dimension. The spectrogram representation is fed into MS-Conv blocks to extract high-level spatial features as the input features of network models.

The proposed MS-Conv block utilizes two dual convolution layers with $3\times 3$ and $5\times 5$ kernels to extract multi-scale features. The multi-scale features are then fused with residual connection using trainable weights ${{w}_{1}}$,${{w}_{2}}$, and ${{w}_{3}}$. The MS-Conv block is formulated as follows:
\begin{flalign}\label{eq8}
&&
    \mathbf{x}' =\operatorname{Dropout}\left( \operatorname{Maxpooling}\left( {{w}_{1}}{{\mathbf{x}}_{\mathbf{-}}}\mathbf{3}+{{w}_{2}}{{\mathbf{x}}_{\mathbf{-}}}\mathbf{5}+{{w}_{3}}\mathbf{x} \right) \right)	
&&
\end{flalign}
where $\mathbf{x}$ and $\mathbf{x}' $ are the input and output of the MS-Conv block, respectively, ${{\mathbf{x}}_{\mathbf{-}}}\mathbf{3}$ and ${{\mathbf{x}}_{\mathbf{-}}}\mathbf{5}$ are the outputs of two dual convolutions with $3\times 3$ and $5\times 5$ kernels. The Swish activation function \cite{ref43} is employed between the dual convolution layers to enhance model accuracy and convergence speed.

\subsubsection{SwG-former model}
SwG-former model first processes the input with four layers of MS-Conv blocks and five layers of SwG-former blocks. MS-Conv block utilizes trainable weights to fuse multi-scale features to extract high-level spatial features. The data flow is sustained at 250 frames to capture abundant sequence information and then condensed to 50 frames through max pooling operation. Subsequently, the FC layers with the tanh function map the output of SELD onto a unit circle in the ACCDOA format. Further details are illustrated in Fig. \ref{FIG:fig5}a.



\subsubsection{SwG-EINV2 model}
\begin{figure} [ht] 
    \centering
    \includegraphics[width=1\textwidth]{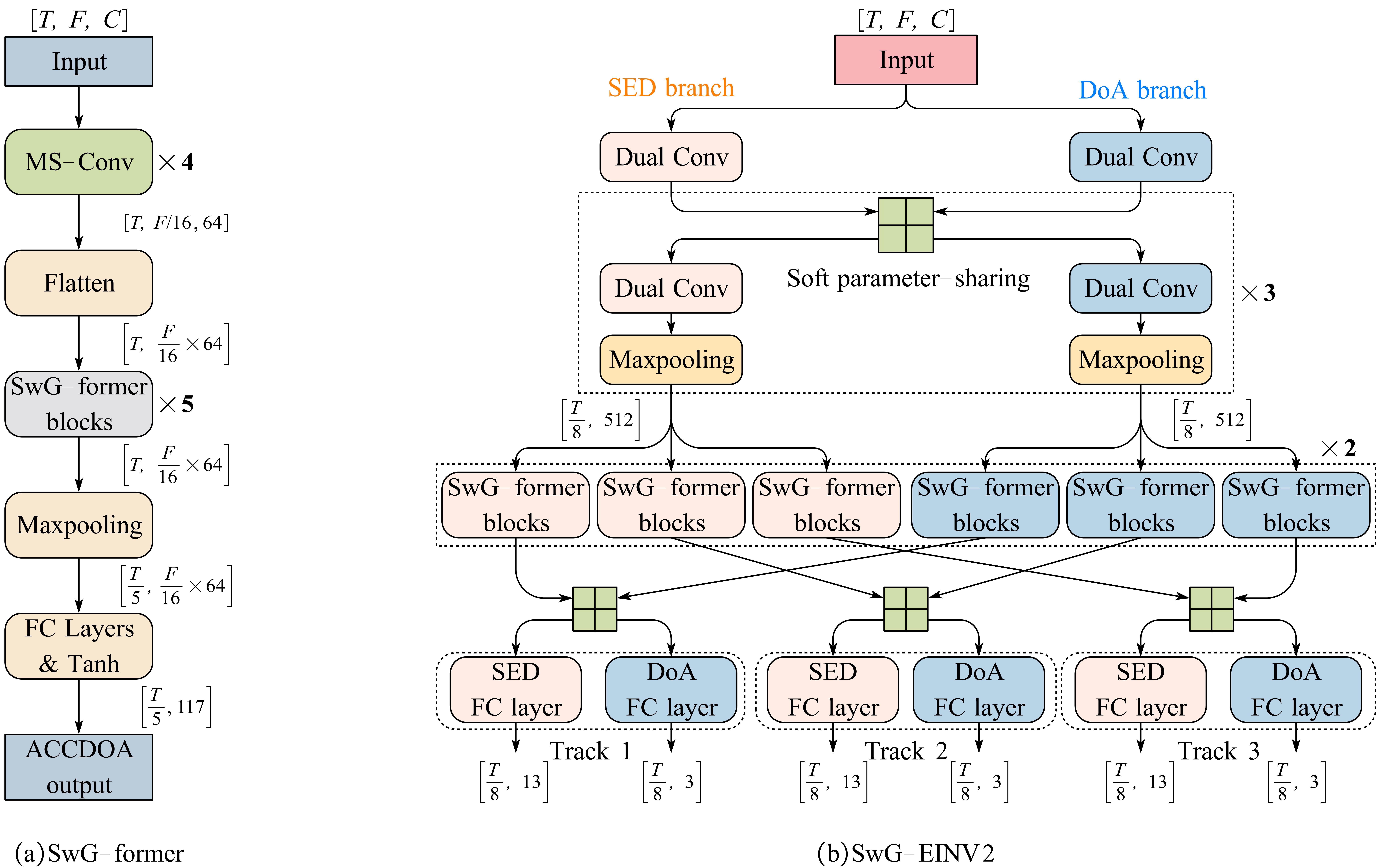}

    \caption{The architectures of the SwG-former and SwG-EINV2 models.}
  \label{FIG:fig5}
\end{figure}

SwG-EINV2 model employs SwG-former blocks to replace the Conformer ones within the EINV2 framework \cite{ref11}. The pink blocks signify the SED task, whereas the blue blocks represent the DoA estimation task. Four-channel log-mel spectrograms are input into the SED branch, while the DoA branch receives both 4-channel log-mel spectrograms and 3-channel IVs. Four layers of Dual Conv (Fig. \ref{FIG:fig4}) process the input and then condense time frames to reduce the model parameters. The green boxes denote the soft parameter-sharing between the SED and DoA subtasks, meaning different subtasks utilize their respective feature layers instead of the same ones. Finally, SwG-EINV2 employs the track-wise output format to manage up to three overlapping sound events, as illustrated in Fig. \ref{FIG:fig5}b.

\section{Experiments}
\subsection{Dataset}
The DCASE 2021 and prior challenges emulate spatial and acoustic properties of soundscapes under realistic conditions. However, the synthetic dataset overlooks the natural temporal occurrences or co-occurrences of sounds within a real scene and the spatial constraints and connections inherent to such scenes. The Sony-TAu Realistic Soundscapes 2022 (STARSS22) dataset \cite{ref44} is captured on natural sound scenes. This real dataset presents significant challenges in real-world acoustic environments, such as interference, overlapping sources, moving sources, and reverberation. Within these real recordings, which range from 30 seconds to 6 minutes, it is common to encounter multiple overlapping sound events with up to 5 overlapping sources. The labels are annotated both temporally and spatially. Both STARSS23 and STARSS22 represent real recordings within identical acoustic environments. However, STARSS22 offers a more significant number of advanced models conducive to comparative analysis.

This study utilizes the STARSS22 dataset as the training base for the proposed models. To address the challenges posed by the scarcity of real data, the study further augments the training data by incorporating 1200 one-minute audio samples from the synthetic dataset \cite{ref45}. 

\subsection{ Evaluation metrics}
The evaluation of SELD adopts the joint evaluation method \cite{ref46}. The first two metrics for SED are the error rate ($ER$) and F-score (${{F}_{20{}^\circ }}$), which are location-dependent, that is to say, considering true positive ($TP$) only under a distance threshold $\text{T}{}^\circ $ from the prediction to the reference. The threshold $\text{T}{}^\circ $ is $20{}^\circ $default. The ${{F}_{20{}^\circ }}$ is a measure of retrieval effectiveness. $ER$ and ${{F}_{20{}^\circ }}$ are formulated as follows: 

\begin{flalign}
&&
    ER=\frac{D+I+S}{N}, F_{20^{\circ}}=\frac{2PR}{P+R},
&&
\end{flalign}

\begin{flalign}
&&
    P=\frac{TP}{TP+FP},\text{ }R=\frac{TP}{TP+FN},
&&
\end{flalign}
where $TP, FP$, and $FN$ denote the true positive, false positive, and false negative, respectively. $D, I$, and $S$ represent the deletion, insertion, and substitution errors, respectively. $P$ and $R$ are the precision and recall metrics, and $N$ is the number of reference sound events, respectively.

For DOA, the localization error ($LE$) and localization recall ($LR_{CD}$) are class-dependent, computed only across each class rather than all outputs. $LE$ is the mean angular localization error between predicted DOAs and their closest reference DOAs. $LR_{CD}$ is a recall metric without any spatial threshold and represents how successfully the system detects localization.
For each class $c\in \left[ 1,...,C \right]$ in a time frame, the distance matrix $\mathbf{D}_c$  is obtained by computing the distance between references and predictions. Considering that multiple instances with the same class label exist simultaneously in the scene, the Hungarian algorithm $\mathrm{H}\left( \cdot \right) $ \cite{ref47} associates the responding predictions and references. Thus, a binary association matrix $\mathbf{A}_{c}^{\left( l \right)}=\mathrm{H}\left( \mathbf{D}_{c}^{\left( l \right)} \right) $ and $l$ represents the time frame. For each class $c$, the $LE_{c}^{\left( l \right)}$ and $L{{R}_{c}}$ are formulated as: 
\begin{flalign}
&&
    LE_{c}^{\left( l \right)}=\frac{||\mathbf{A}_{c}^{(l)}\odot \mathbf{D}_{c}^{\left( l \right)}|{{|}_{1}}}{||\mathbf{A}_{c}^{(l)}|{{|}_{1}}}
&&
\end{flalign}

\begin{flalign}
&&
    L{{R}_{c}}=\frac{\sum\nolimits_{l}{||\mathbf{A}_{c}^{(l)}|{{|}_{1}}}}{\sum\nolimits_{l}{N_{c}^{(l)}}}
&&
\end{flalign}
where $\odot $ denotes element-wise product, $||\cdot |{{|}_{1}}$ is L1-norm. The dimension of $\mathbf{D}_{c}^{\left( l \right)}$ is $M_{c}^{(l)}\times N_{c}^{(l)}$ where $M_{c}^{(l)}$ and $N_{c}^{(l)}$ represent the numbers of $c$-th predicted and reference events in the l time frame, respectively.

The overall $LE$ and $LR_{CD}$ could be formulated as follows:
\begin{flalign}
&&
    LE=\frac{1}{CL}\sum\limits_{c}{\sum\limits_{l}{LE_{c}^{\left( l \right)}}}
&&
\end{flalign}
\begin{flalign}
&&
    L{{R}_{CD}}=\frac{1}{C}\sum\limits_{c}{L{{R}_{c}}}
&&
\end{flalign}

The SELD score is adopted to aggregate all four metrics and formulated as follows:
\begin{flalign}
&&
    SEL{{D}_{score}}=\frac{ER+\left( 1-{{F}_{20{}^\circ }} \right)+\frac{LE}{180}+\left( 1-L{{R}_{CD}} \right)}{4}
&&
\end{flalign}

\subsection{Experimental setup}
The FOA format dataset is utilized in this study, and 7-channel log-mel with IV features are employed as frame-wise features extracted from the original audio data with a sample rate of 24 kHz. Further details can be found in Table \ref{tab:table1}. To simplify the representation of the sizes of sliding-windows across different layers, they are designated in Table \ref{tab:table2}.

The experimental investigations utilize the Ubuntu 20.04 operating system, with 32.00 GB of RAM, an AMD R7950X CPU, and an NVIDIA GeForce RTX 4090 GPU. All models are implemented using the Pytorch framework.

\begin{table}[ht] 
\renewcommand{\arraystretch}{1.3}
 \caption{Training hyper-parameters settings. }
  \centering
  \setlength{\tabcolsep}{8mm}{
  \begin{tabular}{c|cc}
    \toprule
    Models        & SwG-former      & SwG-EINV2   \\ 
    \midrule
    Epochs        & \multicolumn{2}{c}{100}       \\
    Batch size    & \multicolumn{2}{c}{32}        \\
    Pool size     & \multicolumn{2}{c}{(1, 2)}    \\
    Dropout rate  & \multicolumn{2}{c}{0.05}      \\
    \textit{k}    & \multicolumn{2}{c}{24}        \\
    Aggregation   & \multicolumn{2}{c}{Conv2dAgg} \\
    Optimizer     & Adam            & Adamw       \\
    Learning rate & 0.0001          & 0.0003      \\
    Windows size  & group B         & group H     \\
    Layers of GCN & 5               & 2           \\ 
    \bottomrule
  \end{tabular}}
  \label{tab:table1}
\end{table}

\begin{table}[ ht] 
\renewcommand{\arraystretch}{1.3}
 \caption{The representation of different-sized sliding-windows.}
  \centering

  \begin{tabular}{cccc}
    \toprule
    {[}5, 25, 25, 25, 25{]} & {[}5, 5, 25, 25, 25{]} & {[}5, 5, 5, 25, 25{]}    & {[}25, 25, 25, 5, 5{]} \\
    group A             & group B            & group C              & group D            \\
    \midrule
    {[}1, 1, 1, 1, 1{]}     & {[}5, 5, 5, 5, 5{]}    & {[}25, 25, 25, 25, 25{]} & {[}5, 5{]}          \\
    group E             & group F            & group G              & group H            \\
    \bottomrule
  \end{tabular}
  \label{tab:table2}
\end{table}

\subsection{Separate vs. Simultaneous extraction of spatial-temporal information}
This section discusses the earlier concept of simultaneously extracting spatial-temporal information for enhanced model performance, which is the impetus for integrating GCN into the SELD task. These experiments change the size of sliding-windows to the range of input feature frames processed by the SwG-former block in each layer. When the sliding-window size in each network layer is set as one, that setting of group E, the SwG-former block ceases to model context-dependence on temporal information. It exclusively extracts spatial features in the F-C domain. 

In contrast, the SwG-former block, with the settings of groups F, G, and B, simultaneously extracts spatial-temporal information. Groups F and G design the sliding-window sizes in each network layer as 5 and 25, respectively. This design highlights a comprehensive utilization of small and large windows for extracting instantaneous and long-duration temporal information, respectively. Group B represents a compromised choice that balances groups F and G. 

Table \ref{tab:table3} shows the best performance of group B across the multiple experiments, demonstrating a superior SELD score of 0.416. Notably, the simultaneous groups generally outperform the separate group regarding error rate, F-score, and localization error but slightly underperform in localization recall. This can be primarily attributed to the SED task focusing more on the temporal information of audio signals. The simultaneous groups with larger windows can extract sequence context dependencies. Conversely, the DoA task places a higher emphasis on the spatial information inherent in audio signals. As such, with its ability to capture more granular spatial features, the separate group yields a marginal enhancement in localization recall.  
However, the overall performance of group G is somewhat inferior to that of the separate group. Group G employs large windows in every layer, which struggles to capture more granular temporal information. For instance, large windows may simultaneously process multiple classes of sound events, failing to accurately detect and locate instantaneous sound events. Groups B and F alleviate this drawback through smaller window sizes, with Group B achieving a more comprehensive performance through a reasonable window size.

\begin{table}[ht] 
\renewcommand{\arraystretch}{1.3}
 \caption{The performance of separate and simultaneous extraction in SwG-former model. }
  \centering
  \begin{tabular}{cccccc}
    \toprule

     & $\mathrm {ER}\downarrow$ & $\mathrm{{F}_{20{}^\circ }}\uparrow$
    & $\mathrm {LE} \downarrow$ & $\mathrm{LR}_{\mathrm{CD}}\uparrow$ & $\mathrm{SELD}_{\mathrm{score}}\downarrow$ \\
    \midrule
    Group E (separate)     & 0.67                   & 41.8                                                                               & 26.0                   & \textbf{66.1}                       & 0.432                                      \\
    Group B (simultaneous) & \textbf{0.64}          & \textbf{45.2}                                                                      & 24.5                   & 65.7                                & \textbf{0.416}                             \\
    Group F (simultaneous) & 0.65                   & 43.7                                                                               & 23.9                   & 64.6                                & 0.425                                      \\
    Group G (simultaneous) & 0.70                   & 41.5                                                                               & \textbf{23.5}          & 64.1                                & 0.440           \\
    \bottomrule
  \end{tabular}
  \label{tab:table3}
\end{table}

\begin{figure}  [ht] 
    \centering
    \includegraphics[width=1\columnwidth]{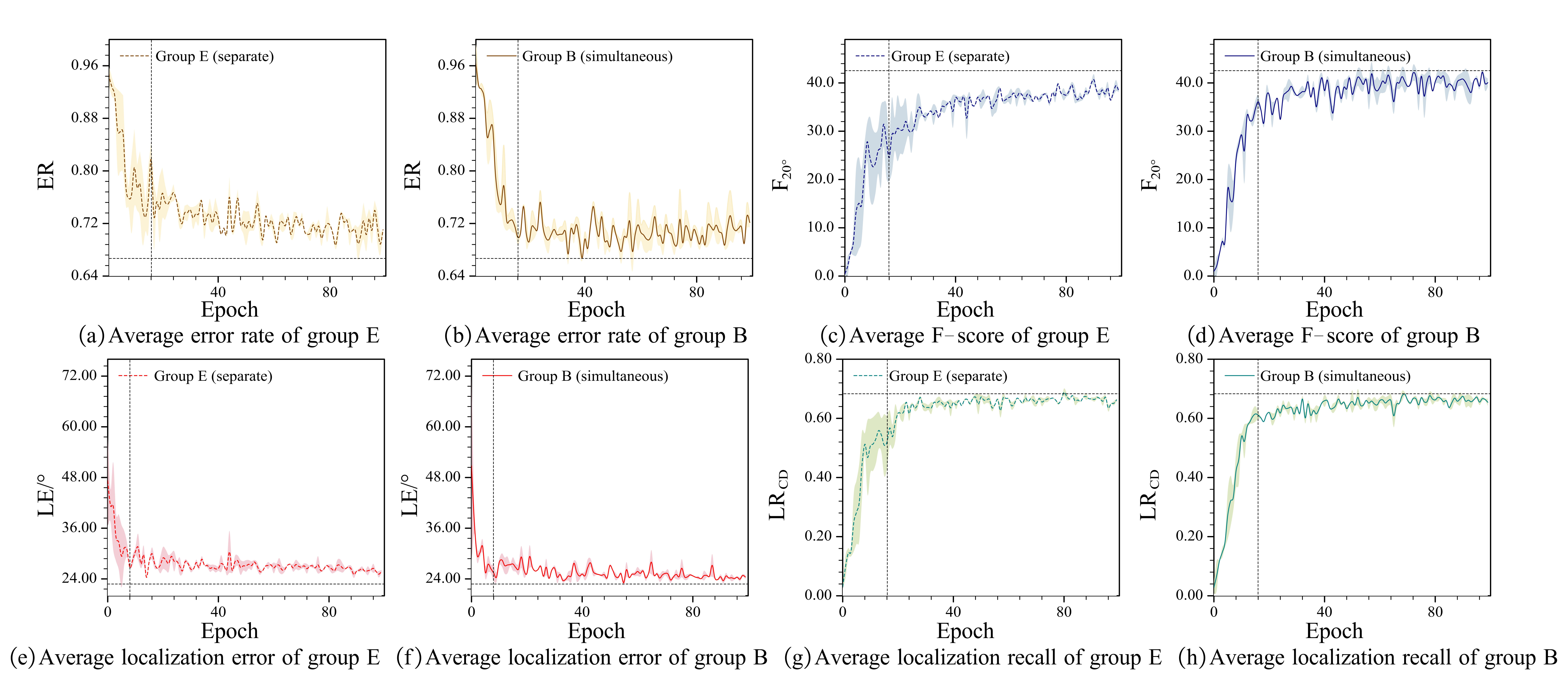}
    \caption{Learning curves of separate and simultaneous extraction in SwG-former model.}
  \label{fig:fig7}
\end{figure}

To understand how group B significantly outperforms group E, Fig. \ref{fig:fig7} presents their average learning curves, evaluated across four metrics. Group B demonstrates a more rapid convergence than group E across all metrics and exhibits more minor fluctuations in its learning curves. As shown in Figs. \ref{fig:fig7} (a), (b), (e), and (f), group B consistently maintains a lower error rate and localization error. Its minimum values significantly outperform those of group E. Regarding F-score and localization recall, group B exhibits superior convergence speed and stability within the initial 20 epochs. That can be attributable to the large window size extracting temporal features. As illustrated in Figs. \ref{fig:fig7} (c), (d), (g), and (h), the F-score of group B, both overall and at its peak, significantly surpasses that of group E. At the same time, the performance of localization recall is relatively comparable between the two groups. This suggests that the simultaneous groups with a reasonable choice of window size could outperform the separate group in SwG-former.

\subsection{Ablation studies}
A series of ablation studies are undertaken to investigate SwG-former performance in this section. The connection strategies of modules initially determine the overall framework of the models. Subsequently, the resolution of feature processing is defined by the size of the sliding windows. Finally, the capacity of message passing among vertices is enhanced by adjusting the aggregation functions and the number of neighbors.

\subsubsection{ Different connection strategies}
A series of ablation studies within the Conformer have substantiated the exemplary performance exhibited by a pair of FF modules surrounding the block in a macaron-style \cite{ref12}. The experiments progressively explored the performance of different module connection orders in the SwG-former block structure. Table \ref{tab:table4} demonstrates that the structure with a pair of FF modules sandwiching the MHSA and SwG modules exhibits the best SELD score performance. By comparing the results from the [FF, MHSA, SwG, FF] structure, the [FF, MHSA, SwG, FF] leads to significantly better results in decreasing SELD scores from 0.484 to 0.416. The result is consistent with Conformer studies wherein macaron-style FF modules are used.

\begin{table}[ht] 
\renewcommand{\arraystretch}{1.3}
 \caption{The performance of different strategies in SwG-former model. [FF, MHSA, SwG, FF] denotes the SwG-former block architecture that is sequentially composed of FF module, MHSA module, SwG module, and another layer of FF module.}
  \centering
  \begin{tabular}{cccccc}
    \toprule

     & $\mathrm {ER}\downarrow$ & $\mathrm{{F}_{20{}^\circ }}\uparrow$
    & $\mathrm {LE} \downarrow$ & $\mathrm{LR}_{\mathrm{CD}}\uparrow$ & $\mathrm{SELD}_{\mathrm{score}}\downarrow$ \\
    \midrule
{[}FF, MHSA, SwG, FF{]} & \textbf{0.64}          & \textbf{45.2}                                                                      & \textbf{24.5}          & 65.7                                & \textbf{0.416}                                                 \\
{[}FF, SwG, MHSA, FF{]} & 0.73                   & 37.6                                                                               & 25.7                   & \textbf{68.4}                       & 0.453                                                          \\
{[}FF, MHSA, FF, SwG{]} & 0.67                   & 39.5                                                                               & 25.6                   & 64.5                                & 0.443                                                          \\
{[}MHSA, FF, SwG, FF{]} & 0.76                   & 34.9                                                                               & 28.1                   & 63.3                                & 0.484                  \\
    \bottomrule
  \end{tabular}
  \label{tab:table4}
\end{table}

\subsubsection{ Size of sliding-windows}
The size of the sliding-windows significantly impacts the resolution of feature processing. A window size that is too small brings more iterative processing, thereby imposing a substantial computational burden. Conversely, a window size is too large to extract local features from instantaneous sound events. Given that the input features comprise 250 frames, window sizes of 5 and 25 were ultimately selected.

Various sizes of sliding-windows were tested in the SwG-former block. Group B demonstrates the best performance in Table \ref{tab:table5}. Comparing the results from group B and group D suggests that using smaller windows followed by larger windows leads to outperformance.

\begin{table}[ ht] 
\renewcommand{\arraystretch}{1.3}
 \caption{The performance of different size of sliding-windows in SwG-former model.}
  \centering
  \begin{tabular}{cccccc}
    \toprule

     & $\mathrm {ER}\downarrow$ & $\mathrm{{F}_{20{}^\circ }}\uparrow$
    & $\mathrm {LE} \downarrow$ & $\mathrm{LR}_{\mathrm{CD}}\uparrow$ & $\mathrm{SELD}_{\mathrm{score}}\downarrow$ \\
    \midrule
Group A & 0.69                   & 41.1                                                                               & 25.0                   & 65.8                                & 0.439                                      \\
Group B & \textbf{0.67}          & \textbf{42.9}                                                                      & \textbf{24.0}          & \textbf{66.7}                       & \textbf{0.427}                             \\
Group C & 0.69                   & 40.4                                                                               & 25.1                   & 63.1                                & 0.449                                      \\
Group D & 0.71                   & 39.4                                                                               & 25.6                   & 65.1                                & 0.451                   \\
    \bottomrule
  \end{tabular}
  \label{tab:table5}
\end{table}

\subsubsection{Type of aggregation functions}

The experiments aimed to evaluate the performance of different aggregation functions, such as MR \cite{ref40}, SAGE \cite{ref37}, GIN \cite{ref39}, and the proposed Conv2dAgg. Table \ref{tab:table6} illustrates the superior performance of the proposed Conv2dAgg, particularly in terms of $ER$ and ${{F}_{20{}^\circ }}$, where it significantly outperforms other aggregation functions. Moreover, Conv2dAgg achieves the second-best performance in terms of $LE$ and $LR_{CD}$. From the SELD score result, it is clear that the Conv2dAgg has a robust fitting capability.

\begin{table}[ht] 
\renewcommand{\arraystretch}{1.3}
 \caption{The performance of different aggregation functions in SwG-former model.}
  \centering
  \begin{tabular}{cccccc}
    \toprule

     & $\mathrm {ER}\downarrow$ & $\mathrm{{F}_{20{}^\circ }}\uparrow$
    & $\mathrm {LE} \downarrow$ & $\mathrm{LR}_{\mathrm{CD}}\uparrow$ & $\mathrm{SELD}_{\mathrm{score}}\downarrow$ \\
    \midrule
Conv2dAgg & \textbf{0.64}          & \textbf{45.2}                                                                      & 24.5                   & 65.7                                & \textbf{0.416}                             \\
MR \cite{ref40}      & 0.67                   & 42.9                                                                               & \textbf{24.0}          & \textbf{66.7}                       & 0.427                                      \\
SAGE  \cite{ref37}    & 0.70                   & 40.5                                                                               & 25.6                   & 63.2                                & 0.450                                      \\
GIN  \cite{ref39}     & 0.70                   & 39.4                                                                               & 25.2                   & 65.1                                & 0.448                    \\
    \bottomrule
  \end{tabular}
  \label{tab:table6}
\end{table}

\subsubsection{ The Number of Neighbors}
In the process of dynamically constructing the graph, the number of neighbors $k$ for each vertex determines the range of aggregated features, similar to the kernel size in CNN. A sufficient number of neighbors hinders message passing, while an excessive number of neighbors may lead to an over-smoothing problem. The results in Table \ref{tab:table7} demonstrate that SwG-former model performs best when the number of neighbors, $k$, is set to 24.

\begin{table}[ht] 
\renewcommand{\arraystretch}{1.3}
 \caption{The performance of different neighbors $k$ in SwG-former model. }
  \centering
  \begin{tabular}{cccccc}
    \toprule

 \textit{k} & $\mathrm {ER}\downarrow$ & $\mathrm{{F}_{20{}^\circ }}\uparrow$
    & $\mathrm {LE} \downarrow$ & $\mathrm{LR}_{\mathrm{CD}}\uparrow$ & $\mathrm{SELD}_{\mathrm{score}}\downarrow$ \\
    \midrule
18         & 0.68          & 40.2                                                                      & 25.6                   & 66.2                                & 0.440                            \\
21         & 0.72                   & 40.1                                                                               & 25.5          & 65.3                      & 0.451                                      \\
23         & 0.69                   & 39.1                                                                               & 25.6                   & 64.9                                & 0.449                                      \\
24         & \textbf{0.64}               & \textbf{45.2}                                                                               & \textbf{24.5}                   & 65.7                                & \textbf{0.416}                                    \\
25         & 0.68                   & 40.3                                                                               & 27.0                   & 66.5                                & 0.440                                      \\
27         & 0.68                   & 41.6                                                                               & 24.6                   & 63.8                                & 0.442                                      \\
30         & 0.71                   & 40.7                                                                               & 25.0                   & \textbf{68.0}                                & 0.440                     \\
    \bottomrule
  \end{tabular}
  \label{tab:table7}
\end{table}

\subsection{ Comparisons with state-of-the-art methods}
The first five rows of Table \ref{tab:table8} illustrate that the proposed SwG-former model significantly outperforms CRNN \cite{ref16}, E-CRNN \cite{ref17}, CNN-MHSA \cite{ref18} and ResNet-Conformer \cite{ref19} across $ER$, ${{F}_{20{}^\circ }}$ and $LR_{CD}$ metrics. Specifically, the SwG-former model markedly improves upon the baseline model by 17.0 and 11.2 in terms of ${{F}_{20{}^\circ }}$ and $LR_{CD}$. Regarding model parameters, SwG-former models possess smaller parameter sizes than the CNN-MHSA model. Despite this, it exhibits a superior comprehensive performance SELD score of 0.416.

\begin{table}[ht] 
\renewcommand{\arraystretch}{1.3}
 \caption{Comparisons of the proposed models with recent SELD models. }
  \centering
  \begin{tabular}{ccccccc}
    \toprule

 Models & $\mathrm {ER}\downarrow$ & $\mathrm{{F}_{20{}^\circ }}\uparrow$ 
    & $\mathrm {LE} \downarrow$ & $\mathrm{LR}_{\mathrm{CD}}\uparrow$ & $\mathrm{SELD}_{\mathrm{score}}\downarrow$ & Parameters\\
    \midrule
CRNN (baseline) \cite{ref16}  & 0.71                   & 28.2                                                                               & 31.6                   & 55.3                                & 0.513                                      & 0.4M       \\
E-CRNN \cite{ref17}          & 0.69                   & 17.9                                                                               & 28.5                   & 44.5                                & 0.538                                      & 10.0M      \\
CNN-MHSA \cite{ref18}         & 0.67                   & 27.0                                                                               & \textbf{24.4}          & 60.3                                & 0.483                                      & 672.0M     \\
ResNet Conformer \cite{ref19} & 0.71                   & 34.7                                                                               & 26.5                   & 59.3                                & 0.479                                      & 13.1M      \\
SwG-former       & \textbf{0.64}          & \textbf{45.2}                                                                      & 24.5                   & \textbf{65.7}                       & \textbf{0.416}                             & 110.8M     \\
\midrule
Conv-Conformer \cite{ref11}  & 0.65                   & 48.4                                                                               & 21.5                   & 70.4                                & 0.396                                      & 325.4M     \\
SwG-EINV2        & \textbf{0.63}          & \textbf{48.9}                                                                      & \textbf{20.9}          & \textbf{71.8}                       & \textbf{0.385}                             & 288.5M                      \\
    \bottomrule
  \end{tabular}
  \label{tab:table8}
\end{table}

To validate the generality and efficiency of the SwG-former block, it has been seamlessly integrated with the EINV2 framework, named SwG-EINV2. Conv-Conformer \cite{ref11} achieved SOTA performance based on EINV2. The final two rows of Table \ref{tab:table8} demonstrate that the four metrics - $ER$, ${{F}_{20{}^\circ }}$, $LE$, and $LR_{CD}$ - uniformly outperform the Conv-Conformer, with the SELD score reaching 0.385 on the FOA format dataset without data augmentation. Within the EINV2 framework, the structure of the SwG-former block is consistent with the results derived from the ablation studies. SwG-EINV2 only employs two layers of SwG-former blocks, reducing the model parameters compared to Conv-Conformer. Notably, the SwG-former block did not undergo fine-tuning within the EINV2 framework, further underscoring its plug-and-play capability.

\begin{figure}  [ht] 
    \centering
    \includegraphics[width=0.52\columnwidth]{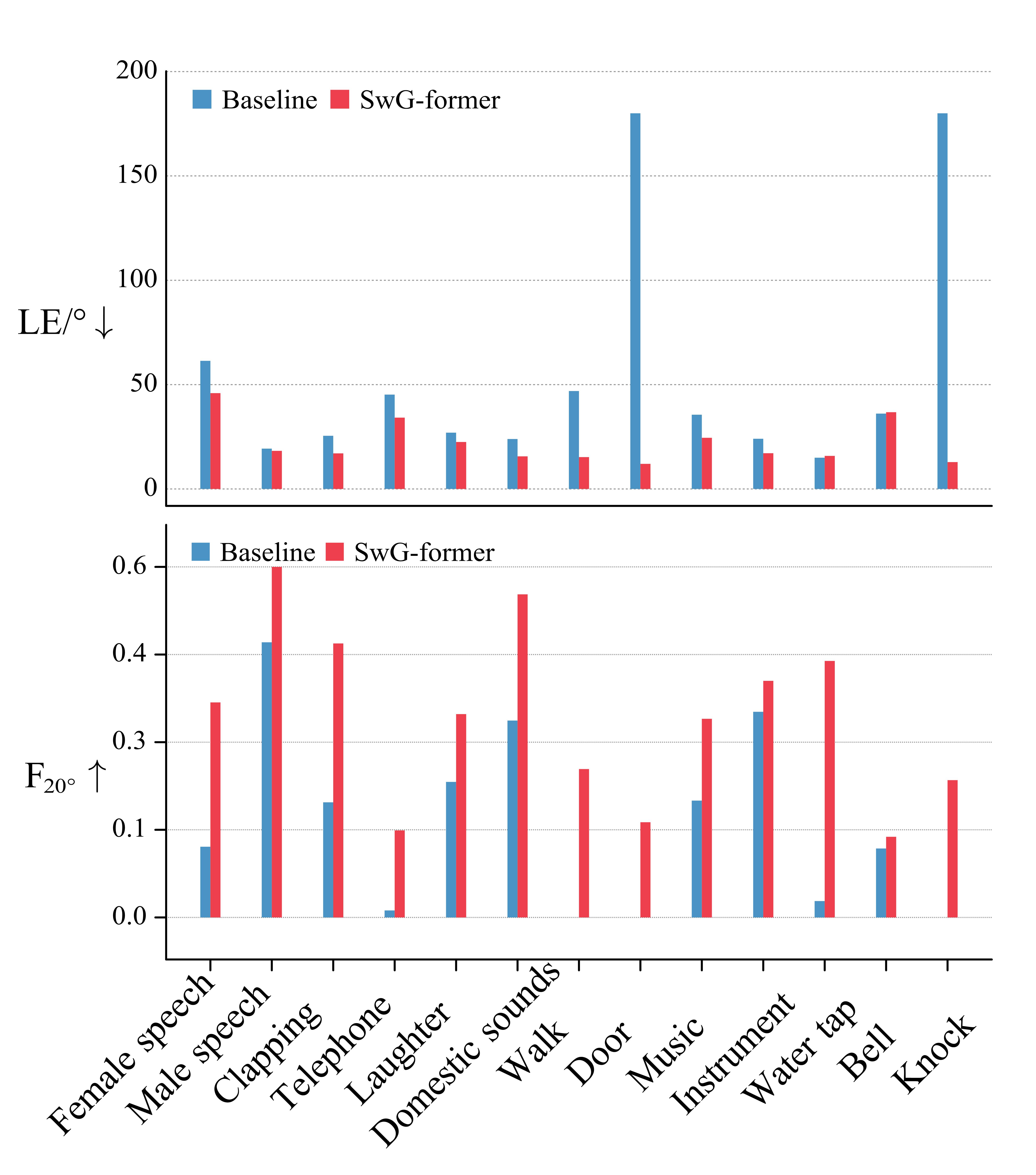}
    \caption{Comparisons of the class-wise performance between baseline and SwG-former models.}
  \label{fig:fig8}
\end{figure}

The class-wise results are illustrated in Fig. \ref{fig:fig8}. The substantial enhancements are discernible in classes such as Telephone, Walk, Door, Water tap, and Knock. In the baseline model, low values in ${{F}_{20{}^\circ }}$, coupled with high $LE$ for Door and Knock, pose significant challenges for the overall metrics. However, in SwG-former model, each class has a noticeable elevation in the ${{F}_{20{}^\circ }}$ and $LE$. Specifically, the $LE$ for Door and Knock has seen a dramatic reduction from 180.00° to 12.04° and 12.88°, while the ${{F}_{20{}^\circ }}$ for Telephone has experienced a significant surge from 0.012 to 0.149. These enhancements across all classes collectively contribute to a superior overall performance.

\subsection{ Visualization results}
\begin{figure} [ht]  
    \centering
    \includegraphics[width=1\columnwidth]{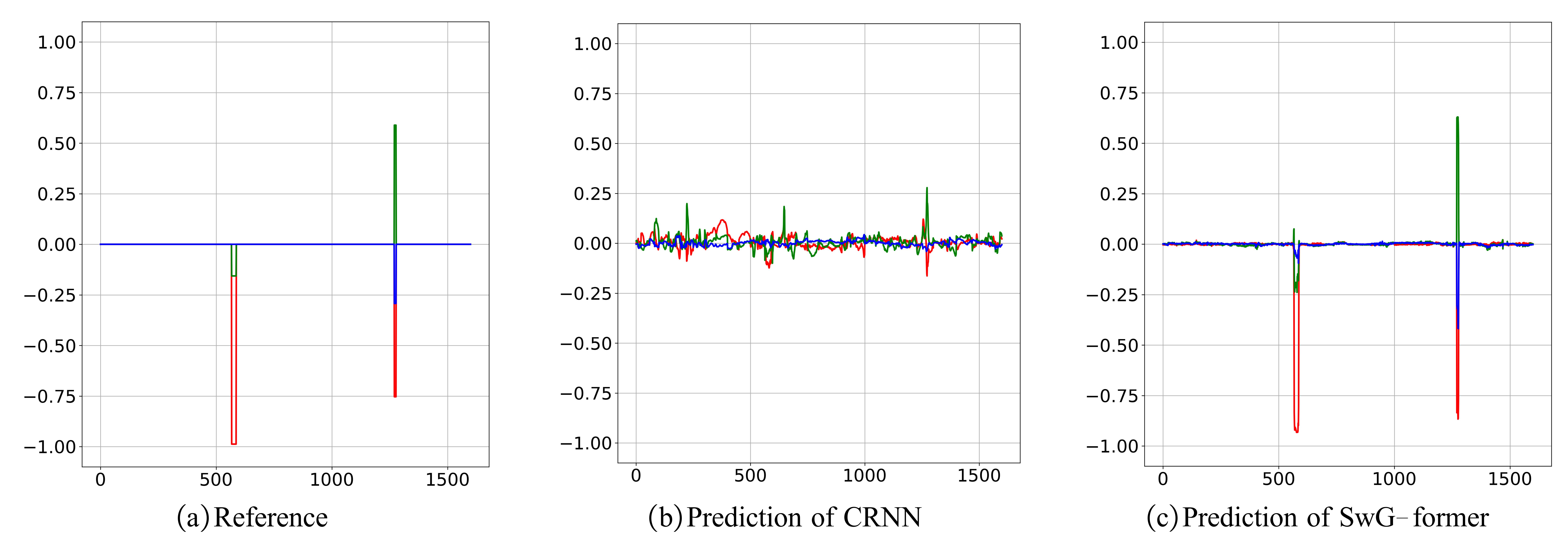}
    \caption{Visualization results for the "Female speech" (Class 0) sound snippets. The abscissa represents the number of frames in the selected audio snippet, and the ordinate corresponds to the corresponding SELD output. The red, green, and blue lines in the figure represent the x, y, and z axes in the Cartesian coordinate system, respectively.   }
  \label{fig:fig9}
\end{figure}

\begin{figure} [ht]  
    \centering
    \includegraphics[width=1\columnwidth]{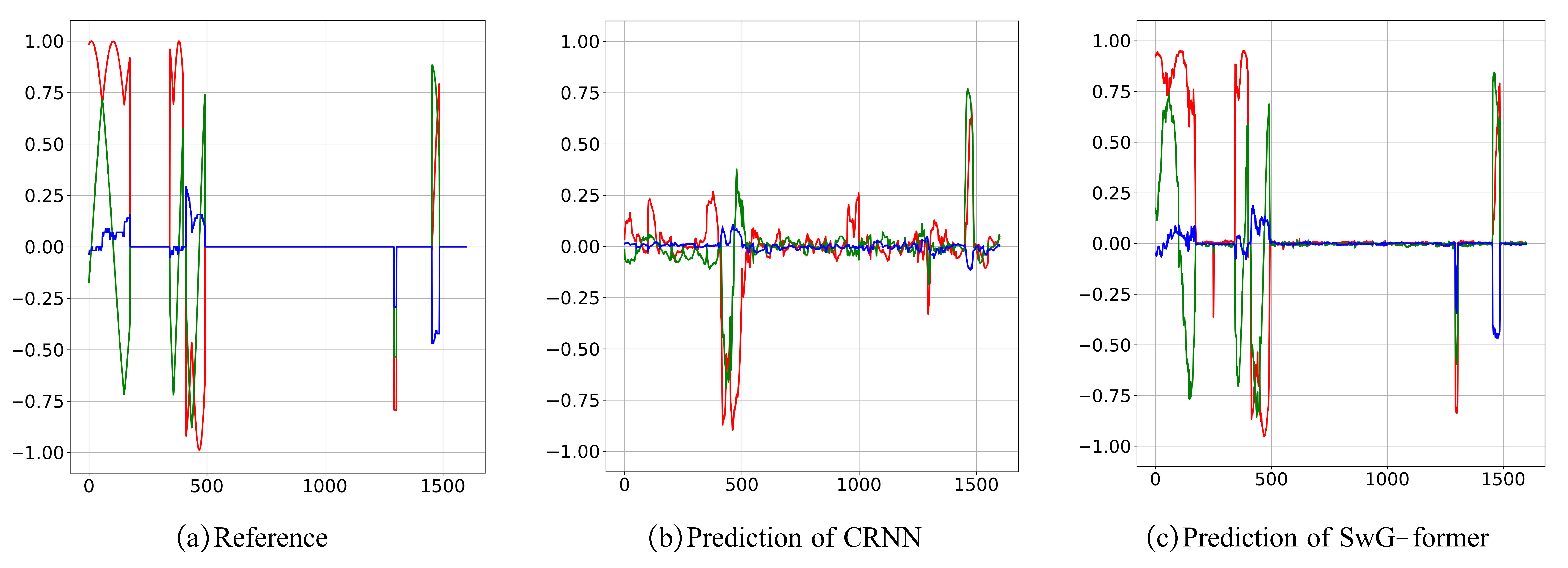}
    \caption{Visualization results for the "Walk" (Class 6) sound snippets. The abscissa represents the number of frames in the selected audio snippet, and the ordinate corresponds to the corresponding SELD output. The red, green, and blue lines in the figure represent the x, y, and z axes in the Cartesian coordinate system, respectively.}
  \label{fig:fig10}
\end{figure}

To more intuitively validate the efficacy of the proposed SwG-former model in practical applications, this study randomly selected several sound event classes with different motion states from the STARSS22 dataset and visually analyzed their sound snippets. The results of this analysis are presented in Figs. \ref{fig:fig9} and \ref{fig:fig10}, all of which originate from the same audio file. Given the relatively broad annotations of the STARSS22 dataset, a specific class label may correspond to multiple different instances. The visible results are divided according to sound event classes. To achieve coordinate normalization, the model mapped the output of SELD to the unit circle using the tanh function. 

Figure \ref{fig:fig9} illustrates the visible results for the instantaneous sound event class "Female speech" (Class 0). Sound events of Class 0 occur around 570 and 1250 frames. It can be observed from Fig. \ref{fig:fig9} (a) that the CRNN model failed to detect the activity of this sound event accurately. However, the comparison results in Figs. \ref{fig:fig9} (a) and (c) demonstrate the effectiveness and superiority of the proposed model.

Figure \ref{fig:fig10} shows the visualization results for the long-duration dynamic sound source "Walk" (Class 6). In the audio, Class 6 appears continuously from 0 to 500 frames, and the sound source location rushes; meanwhile, instantaneous sound sources also appear near 1300 and 1500 frames. It was observed that compared with the CRNN model, the proposed model could more consistently and robustly detect the timing and motion trajectory of the event occurrence under both long-duration and instantaneous states.

\section{Conclusions and future work}
The paper primarily focuses on broadening the application of GCN in SELD tasks. A novel and universal graph representation is proposed to convert audio signals into graph signals, which processes audio more flexibly than the conventional spectrogram representation. Based on the graph representation, the SwG-former block is designed to model local and global context dependencies and extract spatial correlations concurrently by fusing SwG and MHSA modules. As the cornerstone of message passing, a robust 2D convolution aggregation function is proposed and embedded into the SwG-former block called Conv2dAgg. This function aggregates the features of neighbor vertices and enhances the fitting capability of the model. Results from the comparative analysis of simultaneous or separate extraction of spatial-temporal information suggest that the simultaneous extraction with a reasonable choice of window size could outperform the separate extraction in SwG-former. Extensive ablation studies reveal that the proposed SwG-former model significantly outperforms other models on the STARSS22 dataset. To compare with complex models, the SwG-former blocks are integrated with the EINV2 framework, named SwG-EINV2. It achieves an SELD score of 0.385 even in the absence of data augmentation technologies and surpasses the SOTA methods under the same acoustic environment. 

In future work, SwG-former will be applied to the SELD audiovisual system, which takes both audio and video inputs. By utilizing the proposed graph representation, SwG-former will effectively fuse audio and video features into a unified space, thereby facilitating the fusion of cross-modal features. This advanced feature representation approach is expected to enhance the performance of SELD significantly.

\section*{CRediT authorship contribution statement}
 \textbf{Weiming Huang}: Conceptualization, Methodology, Investigation, Formal analysis, Writing – original draft, Data curation,  Writing – review \& editing. \textbf{Qinghua Huang}: Supervision, Funding acquisition,  Writing – review \& editing. \textbf{Liyan Ma}: Writing – review \& editing. \textbf{Chuan Wang}: Investigation, Data curation.
\section*{Declaration of competing interest}
The authors declare that they have no known competing financial interests or personal relationships that could have appeared to influence the work reported in this paper.

\section*{Data availability}
Data will be made available on request. 

\section*{Acknowledgement}
The authors would like to thank the editor and anonymous reviewers for their valuable comments. This work was supported by National Natural Science Foundation of China (61571279).

\bibliographystyle{unsrt}  
\bibliography{biliography}

\begin{thebibliography}{10}

\bibitem{ref1}
Pasquale Foggia, Nicolai Petkov, Alessia Saggese, Nicola Strisciuglio, and Mario Vento.
\newblock Audio surveillance of roads: A system for detecting anomalous sounds.
\newblock {\em IEEE transactions on intelligent transportation systems}, 17(1):279--288, 2015.

\bibitem{ref2}
Weipeng He, Petr Motlicek, and Jean-Marc Odobez.
\newblock Deep neural networks for multiple speaker detection and localization.
\newblock In {\em 2018 IEEE International Conference on Robotics and Automation (ICRA)}, pages 74--79. IEEE, 2018.

\bibitem{ref3}
C{\'e}dric Gervaise, Yvan Simard, Florian Aulanier, and Nathalie Roy.
\newblock Optimizing passive acoustic systems for marine mammal detection and localization: Application to real-time monitoring north atlantic right whales in gulf of st. lawrence.
\newblock {\em Applied Acoustics}, 178:107949, 2021.

\bibitem{ref4}
Pawel Swietojanski, Arnab Ghoshal, and Steve Renals.
\newblock Convolutional neural networks for distant speech recognition.
\newblock {\em IEEE Signal Processing Letters}, 21(9):1120--1124, 2014.

\bibitem{ref5}
Annamaria Mesaros, Toni Heittola, Antti Eronen, and Tuomas Virtanen.
\newblock Acoustic event detection in real life recordings.
\newblock In {\em 2010 18th European signal processing conference}, pages 1267--1271. IEEE, 2010.

\bibitem{ref6}
Tatsuya Komatsu, Yuzo Senda, and Reishi Kondo.
\newblock Acoustic event detection based on non-negative matrix factorization with mixtures of local dictionaries and activation aggregation.
\newblock In {\em 2016 IEEE International Conference on Acoustics, Speech and Signal Processing (ICASSP)}, pages 2259--2263. IEEE, 2016.

\bibitem{ref7}
Chao Pan, Jingdong Chen, and Jacob Benesty.
\newblock Performance study of the mvdr beamformer as a function of the source incidence angle.
\newblock {\em IEEE/ACM Transactions on Audio, Speech, and Language Processing}, 22(1):67--79, 2013.

\bibitem{ref8}
Maurizio Omologo and Piergiorgio Svaizer.
\newblock Acoustic event localization using a crosspower-spectrum phase based technique.
\newblock In {\em Proceedings of ICASSP'94. IEEE International Conference on Acoustics, Speech and Signal Processing}, volume~2, pages II--273. IEEE, 1994.

\bibitem{ref9}
Sharath Adavanne, Archontis Politis, Joonas Nikunen, and Tuomas Virtanen.
\newblock Sound event localization and detection of overlapping sources using convolutional recurrent neural networks.
\newblock {\em IEEE Journal of Selected Topics in Signal Processing}, 13(1):34--48, 2018.

\bibitem{ref10}
Yin Cao, Turab Iqbal, Qiuqiang Kong, Fengyan An, Wenwu Wang, and Mark~D Plumbley.
\newblock An improved event-independent network for polyphonic sound event localization and detection.
\newblock In {\em ICASSP 2021-2021 IEEE International Conference on Acoustics, Speech and Signal Processing (ICASSP)}, pages 885--889. IEEE, 2021.

\bibitem{ref11}
Jinbo Hu, Yin Cao, Ming Wu, Qiuqiang Kong, Feiran Yang, Mark~D Plumbley, and Jun Yang.
\newblock A track-wise ensemble event independent network for polyphonic sound event localization and detection.
\newblock In {\em ICASSP 2022-2022 IEEE International Conference on Acoustics, Speech and Signal Processing (ICASSP)}, pages 9196--9200. IEEE, 2022.

\bibitem{ref12}
Anmol Gulati, James Qin, Chung-Cheng Chiu, Niki Parmar, Yu~Zhang, Jiahui Yu, Wei Han, Shibo Wang, Zhengdong Zhang, Yonghui Wu, and Ruoming Pang.
\newblock {Conformer: Convolution-augmented Transformer for Speech Recognition}.
\newblock In {\em Proc. Interspeech 2020}, pages 5036--5040, 2020.

\bibitem{ref13}
Kazuki Shimada, Yuichiro Koyama, Naoya Takahashi, Shusuke Takahashi, and Yuki Mitsufuji.
\newblock Accdoa: Activity-coupled cartesian direction of arrival representation for sound event localization and detection.
\newblock In {\em ICASSP 2021-2021 IEEE International Conference on Acoustics, Speech and Signal Processing (ICASSP)}, pages 915--919. IEEE, 2021.

\bibitem{ref14}
Ashish Vaswani, Noam Shazeer, Niki Parmar, Jakob Uszkoreit, Llion Jones, Aidan~N Gomez, {\L}ukasz Kaiser, and Illia Polosukhin.
\newblock Attention is all you need.
\newblock {\em Advances in neural information processing systems}, 30, 2017.

\bibitem{ref15}
Eric Guizzo, Christian Marinoni, Marco Pennese, Xinlei Ren, Xiguang Zheng, Chen Zhang, Bruno Masiero, Aurelio Uncini, and Danilo Comminiello.
\newblock L3das22 challenge: Learning 3d audio sources in a real office environment.
\newblock In {\em ICASSP 2022-2022 IEEE International Conference on Acoustics, Speech and Signal Processing (ICASSP)}, pages 9186--9190. IEEE, 2022.

\bibitem{ref16}
Kazuki Shimada, Yuichiro Koyama, Shusuke Takahashi, Naoya Takahashi, Emiru Tsunoo, and Yuki Mitsufuji.
\newblock Multi-accdoa: Localizing and detecting overlapping sounds from the same class with auxiliary duplicating permutation invariant training.
\newblock In {\em ICASSP 2022-2022 IEEE International Conference on Acoustics, Speech and Signal Processing (ICASSP)}, pages 316--320. IEEE, 2022.

\bibitem{ref17}
Shichao Wu, Shouwang Huang, Zicheng Liu, and Jingtai Liu.
\newblock Mlp-mixer enhanced crnn for sound event localization and detection in dcase 2022 task 3.
\newblock Technical report, DCASE2022 Challenge, Tech. Rep, 2022.

\bibitem{ref18}
Yuhao Wang, Yuxin Duan, Pingjie Wang, Yu~Wang, Wei Xue, and Cooperative Medianet~Innovation Center.
\newblock Improving low-resource sound event localization and detection via active learning with domain adaptation.
\newblock Technical report, DCASE2022 Challenge, Tech. Rep, 2022.

\bibitem{ref19}
Qing Wang, Jun Du, Hua-Xin Wu, Jia Pan, Feng Ma, and Chin-Hui Lee.
\newblock A four-stage data augmentation approach to resnet-conformer based acoustic modeling for sound event localization and detection.
\newblock {\em IEEE/ACM Transactions on Audio, Speech, and Language Processing}, 31:1251--1264, 2023.

\bibitem{ref20}
Haoyin Yan, Haitao Xu, Qing Wang, and Jie Zhang.
\newblock The nercslip-ustc system for the l3das23 challenge task2: 3d sound event localization and detection (seld).
\newblock In {\em ICASSP 2023-2023 IEEE International Conference on Acoustics, Speech and Signal Processing (ICASSP)}, pages 1--2. IEEE, 2023.

\bibitem{ref21}
Yusun Shul, Byeong-Yun Ko, and Jung-Woo Choi.
\newblock Divided spectro-temporal attention for sound event localization and detection in real scenes for dcase2023 challenge.
\newblock {\em arXiv preprint arXiv:2306.02591}, 2023.

\bibitem{ref22}
Zonghan Wu, Shirui Pan, Fengwen Chen, Guodong Long, Chengqi Zhang, and S~Yu Philip.
\newblock A comprehensive survey on graph neural networks.
\newblock {\em IEEE transactions on neural networks and learning systems}, 32(1):4--24, 2020.

\bibitem{ref23}
Lei Tang and Huan Liu.
\newblock Relational learning via latent social dimensions.
\newblock In {\em Proceedings of the 15th ACM SIGKDD international conference on Knowledge discovery and data mining}, pages 817--826, 2009.

\bibitem{ref24}
Ling Zhao, Yujiao Song, Chao Zhang, Yu~Liu, Pu~Wang, Tao Lin, Min Deng, and Haifeng Li.
\newblock T-gcn: A temporal graph convolutional network for traffic prediction.
\newblock {\em IEEE transactions on intelligent transportation systems}, 21(9):3848--3858, 2019.

\bibitem{ref25}
Chao Song, Youfang Lin, Shengnan Guo, and Huaiyu Wan.
\newblock Spatial-temporal synchronous graph convolutional networks: A new framework for spatial-temporal network data forecasting.
\newblock In {\em Proceedings of the AAAI conference on artificial intelligence}, volume~34, pages 914--921, 2020.

\bibitem{ref26}
Marinka Zitnik and Jure Leskovec.
\newblock Predicting multicellular function through multi-layer tissue networks.
\newblock {\em Bioinformatics}, 33(14):i190--i198, 2017.

\bibitem{ref27}
Nikil Wale, Ian~A Watson, and George Karypis.
\newblock Comparison of descriptor spaces for chemical compound retrieval and classification.
\newblock {\em Knowledge and Information Systems}, 14:347--375, 2008.

\bibitem{ref28}
Federico Monti, Michael Bronstein, and Xavier Bresson.
\newblock Geometric matrix completion with recurrent multi-graph neural networks.
\newblock {\em Advances in neural information processing systems}, 30, 2017.

\bibitem{ref29}
Joan Bruna, Wojciech Zaremba, Arthur Szlam, and Yann LeCun.
\newblock Spectral networks and locally connected networks on graphs.
\newblock {\em arXiv preprint arXiv:1312.6203}, 2013.

\bibitem{ref30}
Amir Shirian and Tanaya Guha.
\newblock Compact graph architecture for speech emotion recognition.
\newblock In {\em ICASSP 2021-2021 IEEE International Conference on Acoustics, Speech and Signal Processing (ICASSP)}, pages 6284--6288. IEEE, 2021.

\bibitem{ref31}
Panagiotis Tzirakis, Anurag Kumar, and Jacob Donley.
\newblock Multi-channel speech enhancement using graph neural networks.
\newblock In {\em ICASSP 2021-2021 IEEE International Conference on Acoustics, Speech and Signal Processing (ICASSP)}, pages 3415--3419. IEEE, 2021.

\bibitem{ref32}
Tingting Wang, Zexu Pan, Meng Ge, Zhen Yang, and Haizhou Li.
\newblock Time-domain speech separation networks with graph encoding auxiliary.
\newblock {\em IEEE Signal Processing Letters}, 30:110--114, 2023.

\bibitem{ref33}
Charles~R Qi, Hao Su, Kaichun Mo, and Leonidas~J Guibas.
\newblock Pointnet: Deep learning on point sets for 3d classification and segmentation.
\newblock In {\em Proceedings of the IEEE conference on computer vision and pattern recognition}, pages 652--660, 2017.

\bibitem{ref34}
Will Hamilton, Zhitao Ying, and Jure Leskovec.
\newblock Inductive representation learning on large graphs.
\newblock {\em Advances in neural information processing systems}, 30, 2017.

\bibitem{ref35}
Yue Wang, Yongbin Sun, Ziwei Liu, Sanjay~E Sarma, Michael~M Bronstein, and Justin~M Solomon.
\newblock Dynamic graph cnn for learning on point clouds.
\newblock {\em ACM Transactions on Graphics (tog)}, 38(5):1--12, 2019.

\bibitem{ref36}
Thomas~N. Kipf and Max Welling.
\newblock Semi-supervised classification with graph convolutional networks.
\newblock In {\em International Conference on Learning Representations}, 2017.

\bibitem{ref37}
Petar Veli{\v{c}}kovi{\'c}, Guillem Cucurull, Arantxa Casanova, Adriana Romero, Pietro Lio, and Yoshua Bengio.
\newblock Graph attention networks.
\newblock In {\em International Conference on Learning Representations}, 2018.

\bibitem{ref38}
Dzmitry Bahdanau, Kyunghyun Cho, and Yoshua Bengio.
\newblock Neural machine translation by jointly learning to align and translate.
\newblock {\em arXiv preprint arXiv:1409.0473}, 2014.

\bibitem{ref39}
Keyulu Xu, Weihua Hu, Jure Leskovec, and Stefanie Jegelka.
\newblock How powerful are graph neural networks?
\newblock In {\em International Conference on Learning Representations}, 2019.

\bibitem{ref40}
Guohao Li, Matthias Muller, Ali Thabet, and Bernard Ghanem.
\newblock Deepgcns: Can gcns go as deep as cnns?
\newblock In {\em Proceedings of the IEEE/CVF international conference on computer vision}, pages 9267--9276, 2019.

\bibitem{ref41}
Dan Hendrycks and Kevin Gimpel.
\newblock Gaussian error linear units (gelus).
\newblock {\em arXiv preprint arXiv:1606.08415}, 2016.

\bibitem{ref42}
Kai Han, Yunhe Wang, Jianyuan Guo, Yehui Tang, and Enhua Wu.
\newblock Vision gnn: An image is worth graph of nodes.
\newblock {\em Advances in Neural Information Processing Systems}, 35:8291--8303, 2022.

\bibitem{ref43}
Prajit Ramachandran, Barret Zoph, and Quoc~V. Le.
\newblock Searching for activation functions.
\newblock In {\em International Conference on Learning Representations}, 2018.

\bibitem{ref44}
Archontis Politis, Kazuki Shimada, Parthasaarathy Sudarsanam, Sharath Adavanne, Daniel Krause, Yuichiro Koyama, Naoya Takahashi, Shusuke Takahashi, Yuki Mitsufuji, and Tuomas Virtanen.
\newblock Starss22: A dataset of spatial recordings of real scenes with spatiotemporal annotations of sound events.
\newblock {\em arXiv preprint arXiv:2206.01948}, 2022.

\bibitem{ref45}
Eduardo Fonseca, Xavier Favory, Jordi Pons, Frederic Font, and Xavier Serra.
\newblock Fsd50k: an open dataset of human-labeled sound events.
\newblock {\em IEEE/ACM Transactions on Audio, Speech, and Language Processing}, 30:829--852, 2021.

\bibitem{ref46}
Archontis Politis, Annamaria Mesaros, Sharath Adavanne, Toni Heittola, and Tuomas Virtanen.
\newblock Overview and evaluation of sound event localization and detection in dcase 2019.
\newblock {\em IEEE/ACM Transactions on Audio, Speech, and Language Processing}, 29:684--698, 2020.

\bibitem{ref47}
Harold~W Kuhn.
\newblock The hungarian method for the assignment problem.
\newblock {\em Naval research logistics quarterly}, 2(1-2):83--97, 1955.

\end{thebibliography}

\end{document}